\documentclass[doublespacing]{bmcart}
\usepackage{hyperref}
\usepackage{fancyhdr}
\usepackage{subcaption}
\usepackage{graphicx}
\usepackage{enumitem}
\usepackage{listings}
\usepackage{multirow}
\usepackage{verbatim}
\usepackage{amssymb}
\usepackage{xcolor}
\usepackage{amsmath}
\usepackage{calrsfs}
\usepackage{bm}
\usepackage{amsthm,amsmath}
\usepackage[utf8]{inputenc}
\usepackage[textwidth=14cm]{geometry}
\usepackage{chngcntr}
\usepackage{eurosym}
\usepackage{float}
\restylefloat{table}
\usepackage{setspace}
\usepackage{url}
\usepackage{layout}
\usepackage{lineno}

\usepackage{natbib}
\theoremstyle{definition}

\theoremstyle{remark}

\numberwithin{equation}{section}

\renewcommand{\theequation}{\arabic{equation}}

\begin{document}

\begin{frontmatter}

\counterwithout{equation}{section}

\begin{fmbox}

\dochead{Research}

\title{Detecting the sensitive spots in the African interurban transport network}

\author[
   addressref={aff1},
]{\inits{AR}\fnm{Andrew} \snm{Renninger}}
\author[
   addressref={aff1},
]{\inits{VMM}\fnm{Valentina} \snm{Marin}}
\author[
   addressref={aff2, aff1},
]{\inits{CC-A}\fnm{Carmen} \snm{Cabrera-Arnau}}
\author[
   addressref={aff3},
]{\inits{RPC}\fnm{Rafael} \snm{Prieto-Curiel}}

\address[id=aff1]{
  \orgname{Centre for Advanced Spatial Analysis (CASA), University College London}, 
  \street{Gower Stondon},                     %
  \postcode{ WC1E 6AE}                                
  \city{London},                              
  \cny{UK}                                    
}

\address[id=aff2]{
  \orgname{Department of Geography and Planning University of Liverpool}, 
  \street{Chatham St},                     %
  \postcode{L69 7ZT}                                
  \city{Liverpool},                              
  \cny{UK}                                    
}

\address[id=aff3]{
  \orgname{Complexity Science Hub}, 
  \street{Josefstädter Str. 39},                     %
  \postcode{1080}                                
  \city{Vienna},                              
  \cny{Austria}                                    
}


\end{fmbox}
\begin{abstractbox}
\begin{abstract} 

Transport systems are vulnerable to disruption. This is particularly true in Africa, where there are large areas with few highways and heightened risk of violence. Here we attempt to estimate the costs of violent events on African transport in order to understand the way that it may be limiting integration between regions. In the absence of detailed data on trade or migration, we quantify the cost of violence by relating observed incidents to imputed spatial interaction between cities. We produce indices representing the  expected intensity of violent events $\mu$ and the expected strength of interaction $\nu$ between cities in the African interurban network. We estimate the intensity of conflict in a city and, considering the network of all highways on the continent, use a gravity model to generate flows between pairs of cities. We systematically compare $\mu$ to $\nu$ and classify areas according to their combined impact and intensity. Results show that certain cities and roads in the network contain outsize risk to Africa's transportation infrastructure. These cities have a high propensity for subsequent violence against civilians, and given their role in the network, they also substantially effect regional connectivity — and thus economic integration. According to our model, removing just ten edges due to conflict would require rerouting 32\% of trips. The top 100 edges where violence is likely to happen account for 17\% of all trips. We find that cities with the highest $\mu-\nu$ risk are typically small and medium size with large degree, meaning they act as hubs. Vulnerable areas tend to be characterised by the presence of terrorist groups like Boko Haram in Nigeria or Al Shabaab in Somalia.

\end{abstract}

\begin{keyword}
\kwd{conflict; networks; percolation}
\end{keyword}

\end{abstractbox}
\end{frontmatter}

\section{Introduction}

{
Transport systems are vulnerable to congestion, weather, and other events that create delays and disruptions or isolate entire regions \cite{ganin2017resilience, maureira2017everyday, roy2019quantifying}. The costs of disruption are acute in parts of the African interurban transport network for three reasons. First, Africa has a vast and fragmented territory, and the density of its highways is lower than in other parts of the world, so few alternatives exist in the event of network disruptions. Second, deteriorating infrastructure and deferred maintenance make the market for goods on the continent susceptible to increases in transportation costs. Finally, Africa suffers from additional risks not present elsewhere in the world, such as heightened violence in populous regions. It is estimated that between 2001 and 2021, the number of casualties attributed to violence against civilians in Africa increased 260\% \cite{raleigh2010introducing}. Terrorist groups use the fragility of the network to increase their control in some areas and increase the impact of their attacks. In the most extreme cases, groups like Boko Haram and Al Shabaab have taken over entire towns and cities, removing important nodes from the network \cite{chathamTerrorism}. These problems will likely intensify due to extreme weather conditions: a growing body of evidence links droughts and extreme heat with violence, with a standard deviation rise in annual temperatures corresponding to an 11.3\% rise in the rate of intergroup conflict and a standard deviation rise in drought conditions corresponding to a 1.3\% rise in the chance of conflict in an area \cite{burke2015climate, harari2018conflict}.
}

{
Here, we aim to measure the vulnerability of the African transport network using two separate indices: the intensity of events affecting a specific location and the impact that an event has on the estimated flow that travels through the network. 
}

{
We combine data and construct metrics from two sources to understand both the \emph{intensity} of violent events in a location and the \emph{impact} those events have on economic integration. Our first indicator $\mu$ captures the intensity of violent events in an area and our second $\nu$ estimates importance of that area to interurban flows on the continent. To estimate the intensity of violent events, we use data from the Armed Conflict Location \& Event Data project (ACLED) that monitors political violence across Africa, and other parts of the world, mainly from local media reports \cite{raleigh2010introducing}. We construct a self-exciting point process to determine the intensity of events happening at a location based on previous events nearby. Thus, for each node in the network, we construct its daily intensity $\mu_i(t)$. To estimate impact, we use data from Africapolis and the network of all highways on the continent constructed from OpenStreetMap \cite{Africapolis, OpenStreetMap, prieto2022detecting, prieto2022constructing}. Using major highways and roads within the interurban network, it is possible to estimate spatial interaction between any pair of cities and assign it to the shortest route between them based on existing infrastructure. This impact is computed based on the flow that would be observed if a node is removed. For city $i$, we obtain the impact on the network flow $\nu_i$. Note that for both of these metrics, we can also project them along the edges of the network to compute $\mu_{ij}$ and $\nu_{ij}$; in the case of intensity, we simply associate events with the nearest road rather than the nearest city, and in the case of impact we take the total flows along each road when considering the shortest paths between interacting cities.     
}

{
We observe that many nodes in the network have a high likelihood of violence and that such violence would have a high impact on transport infrastructure. We find that the cities with the highest risk tend to be small and medium size with high city degree, that is, cities which create transport shortcuts in the network that otherwise do not exist. Simulating cascading events, we find that cities in West and East Africa are at risk of complete isolation from the network—a finding which comports with historic takeovers by militants in those regions. These areas with the greatest risk of isolation tend to be characterised by terrorist groups like Boko Haram in Nigeria and Al-Shabaab in Somalia.  
}

\section{Literature review}

\subsection{Violence against civilians in Africa is increasing and concentrated} 

{
Data from the Armed Conflict Location \& Event Data project (ACLED) are captured from local media and gives the location, dates, actors, fatalities, and types of reported political violence and protest events worldwide \cite{raleigh2010introducing}. Between 2000 and 2021, ACLED reported over 250,000 events in Africa, with over half a million casualties, including nearly 64,000 events registered as violence against civilians. During 2021, 23 events daily were considered violence against civilians in Africa, causing almost 43 fatalities. This means an increase of more than 400\% in the past ten years. 

Violence against civilians has four key aspects. First, few events account for most casualties \cite{ZipfTerrorism, guo2019common}. Of the 8,000 events classified as violence against civilians in 2021, the top 5\% most violent account for 52\% of the casualties. Thus, roughly one daily event causes half of the civilian casualties in Africa. Second, violence is primarily urban \cite{radil2022urban}. Most violent events happen inside or nearby cities and are suffered directly by their population. Third, violence is highly heterogeneous, meaning that some cities have high levels of violence, whilst others are relatively peaceful \cite{ViolenceNorthWestAfrica, buhaug2008contagion}. For example, Mogadishu (Somalia) had nearly 100 casualties in 2021 related to violence against civilians, whilst Abuja (Nigeria), with a similar population, had only one casualty \cite{raleigh2010introducing}. Finally, violence has some stability over time, meaning that if a city has been at peace for one year, it will likely remain calm for the subsequent years. In contrast, violent cities also tend to stay violent for years. For instance, Mogadishu (Somalia) had nearly 100 casualties in 2021 related to violence against civilians, but it had more than 100 casualties each year for more than a decade, whilst Abuja (Nigeria) had only one casualty in 2021, and it had the same low violence well over a decade. Thus, although violence against civilians is increasing in Africa, it is due to some events highly concentrated in a few cities. As observed in other parts of the world, violence is highly concentrated in a few locations, and they tend to be stable over time, meaning that it is possible to use data from past events to forecast the intensity of future ones \cite{LawCrimeConcentration, CrimeConcentrationVaryingCitySize}. 
}

{
Because violence against civilians is highly concentrated in a few locations and is relatively stable over time, we can model its temporal patterns. The Hawkes process is a mathematical model for a `self-exciting' process \cite{hawkes1971spectra, chuang2019mathematical}. The model is based on a counting process of a sequence of `arrivals' of some event over time, for example, earthquakes, gang violence, trade orders, or bank defaults \cite{MohlerExcitingPointProcess, laub2015hawkes}. Similar models are also used in communication studies where reciprocal information exchange is more likely after a recent message \cite{chowdhary2023temporal}. The idea behind the model is that the probability of subsequent events is high after a major shock. For example, if an earthquake occurs, aftershocks are expected. Similarly, after a recent attack, it is likely that the circumstances that promoted it prevail, so more events are also expected \cite{porter2012self, yuan2019fast, telesca2006global, clauset2013estimating}. We see evidence for this above with the relative persistence of conflict over years. The Hawkes process is a non-Markovian extension of the Poisson process that records recent events. In a Hawkes process, we assume a time-varying intensity $\mu_i(t)$ that depends on all the events in the city $i$ up to time $t$. The Hawkes process produces the intensity function $\mu_i(t)$, which can be interpreted as the ``temperature'' of city $i$. It is a combination of all recent events up to time $t$ that enables us to obtain a daily description of events in city $i$. Furthermore, the function $\mu_i(t)$ enables us to construct some metrics for that city, for example, the per capita intensity.
}

{
Not all events cause the same impact. For example, in 2021, more than 8,000 events were classified as violence against civilians, but five events caused more than 100 civilian casualties, whilst more than 4,000 events had no casualties. Thus, one way to measure the severity of an attack is by the number of deaths \cite{ZipfTerrorism}. Just as an earthquake with a higher seismic intensity is expected to have more aftershocks, we assume that an event with more casualties causes a higher impact and likely more subsequent reactions.
}

\subsection{Urban networks are vulnerable} 

{
Network impact analyses are typically used to observe changes in system behaviour following disruptions modelled by removing network elements (e.g. vertices or edges). The iterative removal process enables the study of the network's overall performance and the propagation of damage within the network by analysing its cascading effects \cite{crucitti2004error, cascadingPorta2004}. Percolation analysis is commonly applied for studying how disconnection occurs in a connected network while elements are progressively removed until a critical threshold is reached, at which the network abruptly breaks into disconnected components \cite{newman2003structure, barabasi2016network}. The phase changes of interconnected components that percolate reveal structural patterns of complex systems, especially concerning the transmission of information within the network \cite{Gallos2011}. Percolation-related processes have also found application in various research domains, such as disease propagation \cite{newman2002spread, Gallos2012}, functional brain networks \citep{Gallos2011}, and the structure of urban systems \cite{ArcautePercolation16, Fluschnik2016, Piovani2017, Behnisch2019, marin2023scalar}. By monitoring the behaviour of the largest connected component, which maintains the communication throughout the system, as well as the isolated components that result after attacks, it is possible to assess the capacity of the network to sustain connectivity under perturbations \cite{Callaway2000}.

Analyses of disruptions in urban spatial networks often study the impacts on street networks and transport systems, which affect communication and exchange between cities. Different studies have examined the effects of shocks on the mobility network and the spatial accessibility between urban centres \cite{gil2008flood}, quantify the variation in the distribution of network centralities \cite{maureira2017everyday}, investigate how street network morphology could affect the transportation properties of the system when facing attacks \cite{PaoloMasucci2016}, and explore the operability of public transport systems under different impacts \cite{berche2009resilience, Derrible2010, pagani2019resilience}. The effects on these spatial networks are often evaluated in terms of the cost required to traverse the network, which can be quantified by measuring the increase in the distance between cities, the volume of rerouted or disrupted flows, or the number of urban entities that remain reachable following a disruption \cite{newman2003structure}. The severity of disruptions and the decline of network performance can vary significantly, influenced by the type and extent of the incident, the connectivity structure of the system, and the role and importance of the disrupted elements \cite{Holme2002}. 
}

{
The location of cities relative to others has relevant socio-economic impacts. Cities near major urban regions can benefit from agglomeration economies \cite{meijers2016borrowing}. Economic activities distribute themselves in response to geographical centrality \cite{StreetNetw}. Proximity, however, does not accurately capture accessibility or travel distance, but details concerning barriers, obstructions, and the existing infrastructure are needed to capture city interactions. Considering different cities as nodes and the road infrastructure between them as the edges of a network is one of the most natural ways to analyse city interactions \cite{barbosa2018human}. Networks provide a structured model and different conceptions and measurements of centrality. 
}

{
When examining a network, there are many ways to measure a node's importance, for example, by its degree or a combination of its degree and the degree of neighbours or other node attributes \cite{buechel2013dynamics, singh2020node}. Node betweenness is a commonly used metric that gives the number of times a node occurs in the shortest paths between all possible pairs of nodes in a network \cite{freeman1978centrality}. However, we use a weighted node betweenness to detect the importance of cities. The weights correspond to a combination of two urban properties: the cities' population and the distance between them. More journeys are expected to happen to and from bigger cities, and fewer journeys are expected between distant locations. Thus, we combine both factors using a gravity model to estimate the number of interurban trips, one of the most widely-used models for considering the strength of interactions between spatial units, in this case, the flow of people. The gravity model generally uses population size and the physical distance, the road distance or the travel time between origin and destination to estimate the number of journeys between two cities \cite{Gravity}. This procedure for obtaining the weighted betweenness based on the gravity model is our method to assess the relevance that a city has within a network. Cities with high intermediacy receive considerable traffic. If traffic through a city with high intermediacy is disrupted, consequences for inter-city travel are high. Therefore, we consider the importance of cities based on two indicators, their size and their positioning within the broader landscape of urban connectivity.
}

{
Here, we aim to decode urban integration by analysing the connectivity of cities and monitoring the impact when parts of the network are removed from the system \cite{ArcautePercolation16, baggag2018resilience}. The removal of cities or roads from the system aims to capture a scenario when journeys are no longer possible due to the presence of a violent group. If violence becomes dominant in a city, trips routed through that city must find an alternative and possibly longer route. Thus some journeys between cities will ultimately be discouraged due to violence. A similar strategy has been applied to quantify how many people lose access to healthcare during crises or natural disasters, finding that some road segments are more important and some locations are more vulnerable regarding access to hospitals \cite{schuster2023stress}.
}

\section{Methods}

{
For the constructed interurban network of African cities and highways, we compute two metrics. One is used to proxy the intensity of violence or risk that a place has of a sizeable disruption. The second metric is used to detect the impact of that disruption in terms of the network flow. Both metrics are constructed at a city level, so we have the intensity of events, $\mu_i(t)$ and the impact $\nu_i$ for city $i$. 
}

\subsection{Measuring the intensity of events}

{
Although there are many expressions for a Hawkes process \cite{laub2015hawkes}, we express the intensity for city $i$ as $\mu_i(t)$ or

\begin{equation} \label{HawkesEqn}
\mu_i(t) =  \sum_{s: t \geq t_s} I_i(s) \kappa_s \exp (-\phi (t-t_s)),
\end{equation}
where $I_i(s) = 1$ if event $s$ is identified as happening in city $i$ and zero otherwise. Equation \ref{HawkesEqn} only considers events that happened before some time $t$ measured in days. The intensity $\mu_i(t)$ increases (jumps) in $\kappa_s>0$ units immediately after the occurrence of an event at time $t_s$, and the intensity decays exponentially with time at rate $\phi>0$. For an event taking place at time $t_s$, we set the parameter $\kappa_s$ to be equal to the reported number of fatalities. Other definitions of $\kappa_s$ could depend on the size of the event (number of protesters, for instance, or number of police officers employed), the state intervention (number of arrests), the social media impact (number of tweets related to the protest), or the media coverage (the space devoted in the newspapers to such event). The parameter $\phi$ is the ``cooling'' parameter of the Hawkes process and refers to the speed at which one event is more or less significant in future days. With high values of $\phi$, the process has a faster memory decay, in which case $\mu_i(t)$ captures the number of casualties of recent days. The number $\mu_i(t)$ is the expected number of casualties in city $i$ at time $t$ considering all events in $[0, t]$. The value of $\phi$ used is the same for all cities assuming a homogeneous cooling-down effect, and it is estimated by maximising the correlation between $\mu_i(t)$ and the observed number of casualties during the two weeks after time $t$. Therefore, the obtained $\mu_i(t)$ gives the best estimate of casualties in the upcoming 14 days. The estimated value of the cooling parameter is $\phi^\star = 0.0987$. More details about the estimation process are in the Supplementary information. The result is $\mu_{i}(t)$ for each city, which we then project to the edges around it with a simple average to create $\mu_{ij}(t)$.
}

\subsection{Estimating the flow between cities}

{
With limited data on trade and migration, we estimate the flow of goods or migrants along each path in the network with a gravity model using available data. This assumes that the flow between cities $i$ and $j$ is a function of the road distance between them and their population. The predicted flow of people between location $i$ and location $j$ is modelled as Equation \ref{GravityEqn}
\begin{equation}
\label{GravityEqn}
F_{ij} = k\frac{P_i \times P_j}{d_{ij}^\beta}
\end{equation}
\noindent
where $F_{ij}$ is the flow, $P_i$ and $P_j$ are the populations of the two locations, respectively, and $d_{ij}$ is the road distance between them. $k$ is used to normalise all flows so they sum to 1, thus conveying a share of all flows rather than the number of travellers using the road. We construct our estimate without fitting to observed data, so we test the sensitivity of our results to various values of $\beta$, including values obtained from the literature on migration in Africa \cite{prieto2022detecting}. For the following experiments, we use $\beta = 2.8$, but checks for robustness are included in the Supplementary information. We limit the maximum travel distance in the model to 3,500 km, equivalent to the distance from Lagos (in Nigeria) to Dakar (in Senegal). This value represents the span of an aspiring economic corridor and an area of proposed economic cooperation \cite{/content/publication/9789264265875-en}—so, in a sense, an optimistic scenario for economic integration. The gravity model gives us flows between cities in the network. Because cities have many road segments between them we then project flows between $i$ and $j$ for all $F_{ij}$ along the shortest path $P_{ij}$ in the African urban network described above before summing the total flows along each edge to give us $\nu_{ij}$. The total flows to and through a city are summed to construct $\nu_{i}$.
}

\subsection{Percolation and violence}

{
We use two approaches to understand the resilience of the interurban network to shocks. The first, attack percolation, samples edges in the network in proportion to conflict intensity $\mu_{ij}(t)$ until we are left with an empty graph, with all cities isolated from others. The remaining strategy, cascading events, assumes that violence is correlated with the potential to spill from one location to another. A Mantel test (a statistical test of the correlation between two matrices) \cite{mantel1967detection} looks at whether the distance between events in time is correlated with the distance between those events in space and uses a simple Pearson correlation between the spatial distance matrix and temporal distance matrix. In our data, the location of violence in space and time is not correlated over the entire period ($r=0.1$, $p<0.05$). When we split the data by country and year, a third of the space-time partitions have $r>0.5$; because the dataset covers a long period correlations do not capture the peaks in violence. Knox tests (a test for space-time interaction using space and time distances) also show that we see far more attacks close in space and time than we would expect to occur by chance \cite{Knox}. We can also look into the data and see that in 2013, militants took control of three towns in Mali: Kidal, Gao and Timbuktu. In 2016, Djibo, Arbinda and Dori in Burkina Faso came under the influence of militant groups. These tests confirm other work, which shows that conflict in one area raises the chance of conflict in adjacent areas \cite{harari2018conflict}.  
}

\paragraph{Percolation.} Here, we remove nodes from the network and observe changes in its properties \cite{ArcautePercolation16}. Rather than removing nodes at thresholds according to distance or risk, we use a stochastic approach. At each step, we sample an individual edge in proportion to conflict intensity $\mu_{ij}(t)$, remove it from the network, and monitor the changes to the network that are triggered by the removal of the edge. We proceed, gradually removing edges according to $\mu_{ij}(t)$ and thus — with some variance — move across the intensity probability distribution. We allow the network to accumulate damage, simulating the effect of prolonged and growing conflict. Although a network spanning the continent will not see broad enough conflict to disrupt the network to the extremes we test here, the patterns shown can reveal aspects of the network structure. 

\paragraph{Cascading events.} We then use a form of bursty percolation. It assumes that violence can spill from one area to another in each round. The difference is that it does not allow damage to accumulate as before: we reset the network each round and run repeated simulations. In this approach, we draw an edge from the network with a probability proportional to its intensity $\mu_{ij}(t)$ and remove it. We then take all edges within 100 km of the seed edge and perform Bernoulli trials for each. In these trials, the probability of failure at proximate edges is weighted by the intensity of conflict associated with each edge. This is akin to performing a biased coin flip, where the level of violence determines the coin's bias. Each edge that ``fails'' is also removed from the network along with the seed edge. At the end of each round, we compute a series of statistics to measure the damage done. We reconstruct the entire network after each round and observe the distributions of these statistics to the network after 1000 rounds, simulating the effect of limited but cascading conflict akin to what has occurred in parts of the continent at various times over the past decades.    

\subsection{Measuring the impact of node removals}

{
At each step of the percolation, we track the population of the largest component as it fragments due to conflict along the network, the population of all isolates created by the attacks on the network, and the percentage of all flows that need to be rerouted to preserve trade and migration on the continent. First, the largest component: by monitoring the size of the largest component ---weighted as its total population--- we look at threats to economic integration on the continent, which requires trade and migration between cities. Second, isolates: a simple measure of the affected population is the total population of all isolates created at each step of the process as we remove edges. Third, affected flows: the percentage of flows that rely on edges removed in the percolation. We also compute a detour (in km) resulting from rerouting flows between a pair of origin and destination nodes, which can no longer take place through the shortest path due to road disruption. For computational reasons, this measure does not use gravity estimates, so it is based strictly on the road network.
}

\section{Results}

{
The intensity for city $i$ of future events $\mu_i(t)$ is estimated by adding the number of casualties. Each event is affected by memory decay. One casualty in the city at time $t_s$ has an immediate effect of adding one unit to the intensity, so $\mu_i(t_s) = \mu_i(t_s-1)+1$. However, as new events happen, the value of past information decreases. We obtain that after one week, one casualty adds only 0.5 units to $\mu_i(t_s +7) = \mu_i(t_s-1)+0.5$. After one month, one casualty at time $t_s$ only adds 0.05 units to the intensity, so $\mu_i(t_s + 30) = \mu_i(t_s-1)+0.05$. Thus, unless more casualties occur in the city, the intensity $\mu_i(t)$ cools down after time $t_s$. For the edge $ij$ we compute its intensity $\mu_{ij}(t)$ as the average of $\mu_{i}(t)$ and $\mu_{j}(t)$.
} 

{

\begin{figure} \centering{
\includegraphics[width=0.6\textwidth]{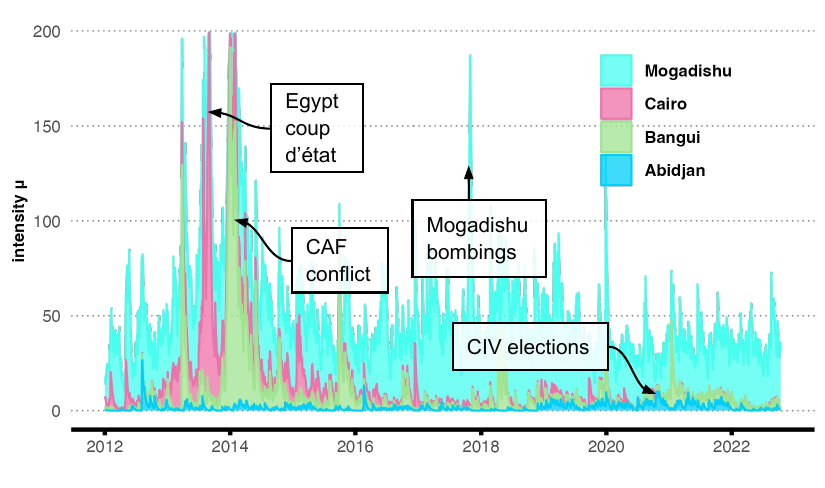}
\caption{Observed values of the intensity of events $\mu_i(t)$ (vertical axis) between 2013 and 2022 in four major cities in Africa, with recurrent violence in Mogadishu and fluctuations elsewhere. } \label{HawkesRes}}
\end{figure}

For some cities, such as Mogadishu (Somalia), the intensity $\mu_i(t)$ is maintained with high levels, indicating an intense and stable violent process. In other cases, the intensity $\mu_i(t)$ reflects specific moments, such as elections or conflict. High values of $\mu_i(t)$ correspond to significant outbreaks of violence in that city. For example, the rate in Cairo (Egypt) had values above 500 in 2013 during Egypt's coup d’état, indicating more than 500 casualties per day (Figure \ref{HawkesRes}). A similar value was observed in Mogadishu (Somalia) in 2015 during its bombings. In Bangui (Central African Republic), roughly 300 casualties were observed daily between 2013 and 2014, corresponding to its conflict. During the 2020 national elections in Ivory Coast, there was also an increase in the daily casualties in its main city, Abidjan, but the daily number of casualties was below ten units during the whole period. The Supplementary information shows how conflict is evolving as violence increases in West Africa, decreases after the Arab Spring in North Africa, and persists in the Horn of Africa.

}

\subsection{Relationship between impact and intensity}

We identify edges and nodes with a high impact disrupting the network $\nu_i$. Edge impact is derived from flows along those roads; for node impact, we recast these weights so that each node is assigned an impact according to the sum of flows to and through it according to the gravity model. Generally, cities with high traffic are characterised by a high centrality, creating shortcuts in the network between big cities. Because of its population, most cities in Egypt have a high impact on the network, as removing them would reduce integration in the country. The same pattern occurs in parts of Nigeria. Cities with a low impact are more isolated, tend to have a smaller degree, or are adjacent to redundant travel options. For example, cities in South Africa tend to have a small impact, as well as cities in Ethiopia. They reflect either high isolation from the network or rich connection to it (Figure \ref{riskvflow}).

{
\begin{figure} 
\centering{
\includegraphics[width=.7\textwidth]{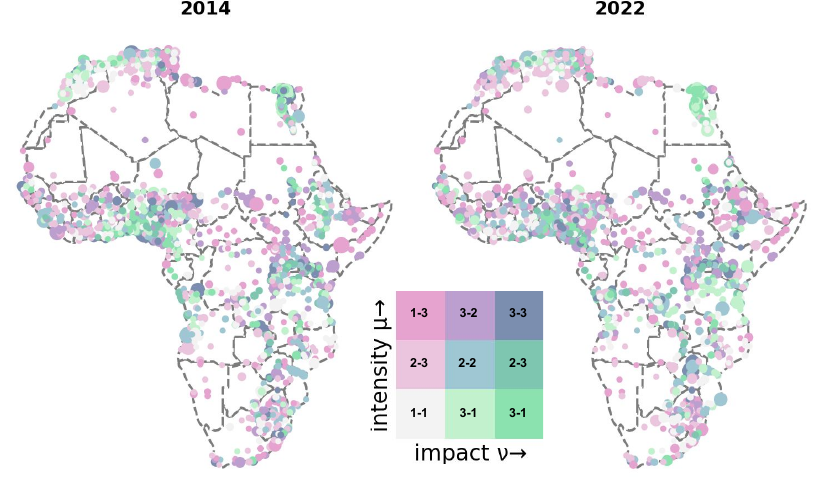}}
\caption{Relationship between conflict intensity and mobility aggregated to cities, with the size of the dot indicating the size of the city, showing the confluence of high intensity and high mobility in West Africa, while in populous areas in North Africa, there is flow between cities but — especially in Egypt in recent years — the absence of violence. Violence in Ethiopia and Nigeria rose between 2014 and 2022.} 
\label{riskvflow}
\end{figure}
}

We analyse the intensity of cities $\mu_i(t)$ for $t = 2022$. That is, we observe the 2022 year and analyse the expected levels of violence for future days. Excluding Libya, which been in civil war, most cities in North Africa have low intensity, reflecting fewer violent outbreaks in the region. However, cities in a band across the Sahel — including those in Nigeria, Mali, Sudan, Somalia or Ethiopia — have higher intensity, reflecting recent episodes of violence (Figure \ref{riskvflow}). By looking at different periods, we can look at how these patterns are changing; we map $\mu_i(t)$ for $t = 2014$ for comparison. Conflict in Nigeria, in particular, is growing. 

Based on the propensity for future events and the impact that an intervention could have, we identify nine groups according to their joint $\mu - \nu$ class (Figure \ref{riskvflow}). The class given to each city is simply the quantiles $\mu$ and $\nu$ assigned to it. Group 1-1 are cities with low impact and low intensity, meaning that due to violence, there are few events expected there, and if some outbreak of violence occurs, the impact will be reduced. In Africa, there are 280 cities with a population of 18 million in this group. In total, cities with the lowest risk of conflict represent 13.5\% of the continent's urban population. We report these results by class and region in Table \ref{RegionContinent}.

\begin{table}[h!]
\centering
\begin{tabular}{|r|cc |cc cc cc cc cc|} 
 \hline
Group & \multicolumn{2}{c|}{Africa} & \multicolumn{2}{c}{North} & \multicolumn{2}{c}{West} & \multicolumn{2}{c}{East} & \multicolumn{2}{c}{Central} & \multicolumn{2}{c|}{South} \\ ($\mu - \nu$)
& cities & pop & cities & pop & cities & pop & cities & pop & cities & pop & cities & pop \\
\hline
1-1 & 280 & 18 & 110 & 8 & 48 & 2 & 34 & 2 & 40 & 2 & 48 & 3 \\
1-2 & 207 & 15 & 85 & 7 & 39 & 2 & 24 & 2 & 18 & 1 & 41 & 3 \\
1-3 & 161 & 23 & 114 & 18 & 17 & 1 & 6 & 1 & 3 & 0 & 21 & 2 \\
2-1 & 181 & 19 & 37 & 3 & 46 & 4 & 41 & 6 & 26 & 3 & 31 & 4 \\
2-2 & 178 & 24 & 42 & 7 & 54 & 7 & 28 & 4 & 15 & 1 & 39 & 5 \\
2-3 & 190 & 81 & 119 & 61 & 24 & 4 & 11 & 2 & 6 & 0 & 30 & 14 \\
3-1 & 118 & 15 & 11 & 4 & 42 & 3 & 10 & 2 & 14 & 2 & 41 & 4 \\
3-2 & 217 & 54 & 17 & 3 & 87 & 22 & 19 & 2 & 13 & 2 & 81 & 25 \\
3-3 & 241 & 164 & 11 & 10 & 121 & 66 & 31 & 31 & 25 & 20 & 53 & 37 \\
\hline 
\end{tabular}
\caption{Number of cities and population (in millions) assigned to separate $\mu-\nu$ groups per continent region. }
\label{RegionContinent}
\end{table}

Many cities in the highest $\mu-\nu$ class have the largest populations. The 241 cities in this group have 164 million inhabitants or 39\% of the total. There is significant violence in West Africa, a populous and busy economic corridor home to Africa's largest economy (Nigeria) and largest city (Lagos), and many cities at risk in East Africa. There are some cities with a high impact $\nu_i$ and a high intensity $\mu_i(t)$, reflecting locations where a violent outbreak is likely to happen and where the impact in the network could be huge. This is the case, for example, for cities in Kenya, such as Nairobi. Although its levels of violence are smaller than elsewhere (including cities in Somalia), its network position creates a higher impact. There are just 23 million people in the low-risk and high-flow — and thus peacefully integrated — condition. 

We examine pairwise correlations between $\mu_{ij}(t)$ and $\nu_{ij}$ along the edges in 2022 to determine if there is an association between conflict and flow on the network. Without a more robust empirical strategy, we cannot decompose cause and effect, but we can explore whether or not relevant corridors are also prone to conflict (Figure \ref{riskyregions}). The strongest correlation is in the South, including a high $\mu - \nu$ along an edge connecting Johannesburg, in South Africa, a path facilitating 16\% of all flows. Many of the highest combined $\mu - \nu$ values are in the West, with a road between Ikorodu and Lagos in Nigeria. This road accounts for 4\% of total flows. Both of these cities are essential hubs in the network and vital centres of activity in the economy.

{
\begin{figure} 
\centering{
\includegraphics[width=0.7\textwidth]{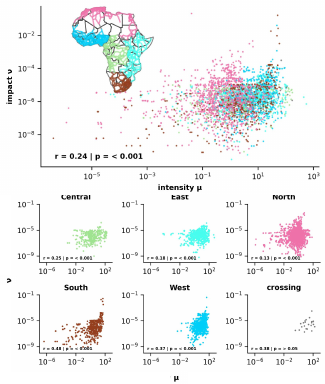}}
\caption{Intensity of conflict $\mu_i(t)$ (horizontal axis) and impact $\nu_i$ (vertical axis) in 2022 according to the region. Areas where violence and mobility interact, are represented by points in the top right quadrant. Critically, many edges in the network that bridge regions have high levels of conflict.}
\label{riskyregions}
\end{figure}
}

We also look at the detour resulting from the removal of an edge. This is the change in the average shortest path on the network after removing that edge (in km). This varies from 1 km to 10,000 km. Most of the largest detours, highlighted in Figure \ref{detours}, cross regions, indicating that continental integration would become more costly if violence occurs in these areas. Aside from crossing roads, the highest detours are in the North and East, while the South and West have redundancy. 

{
\begin{figure} 
\centering{
\includegraphics[width=.7\textwidth]{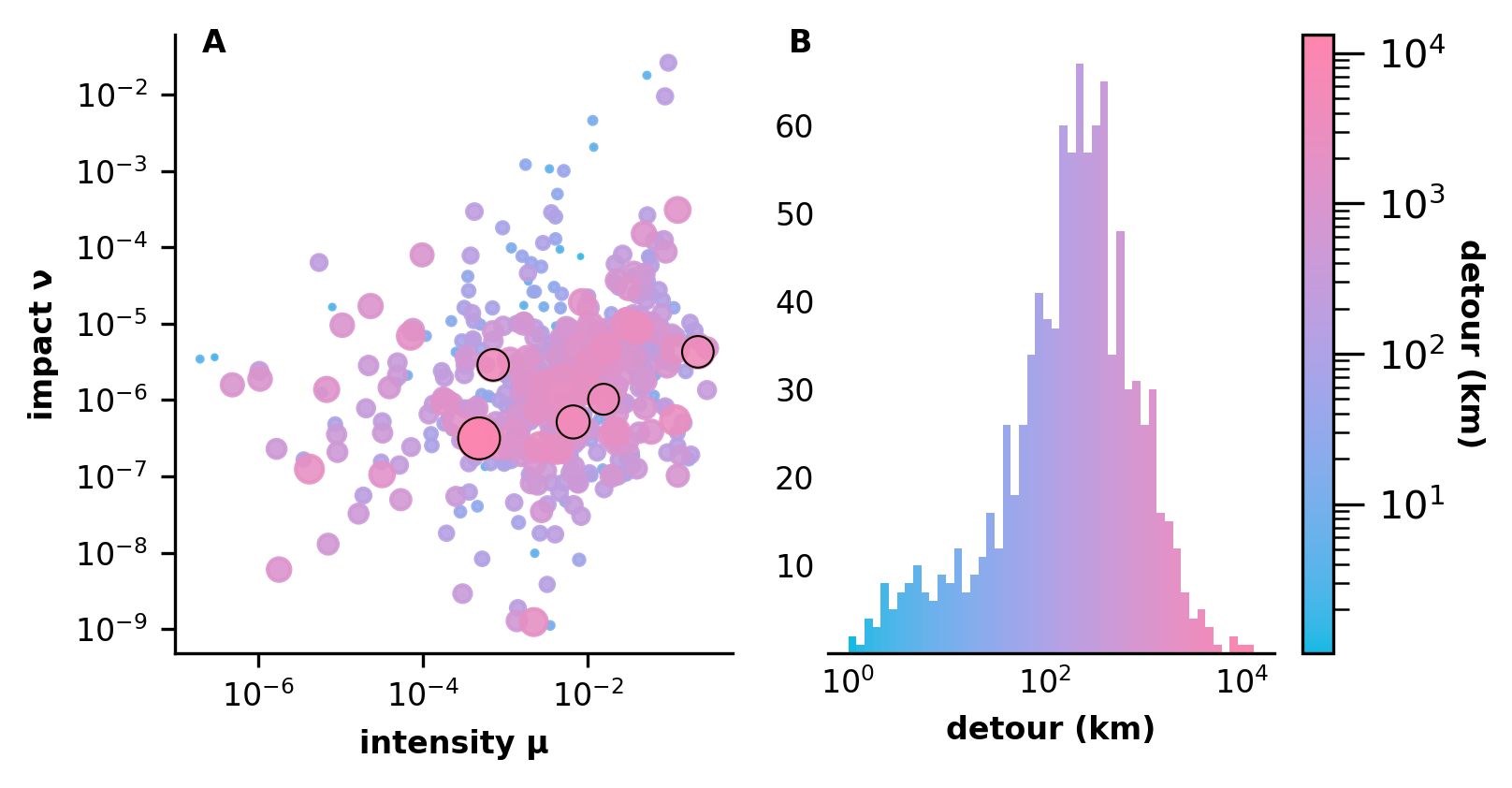}}
\caption{\textbf{A} The relationship between conflict, impact and detours, defined and the change in the average shortest path (km) after removing the edge, along with the distribution of detours. The five largest detours — including the largest two crossing the Sahara — are highlighted. \textbf{B} The distribution of detours: most are just a few hundred or thousand km, but some reach 10 thousand.}
\label{detours}
\end{figure}
}

\subsection{How robust is the system?}

The impact of a network intervention is measured by the number of people that would be removed from the connected component (Figure \ref{simulation1}). With fewer highways, results show that East Africa has the highest vulnerability since a network disruption would disconnect more people. With more highways, many of the most prosperous and populous cities are in North and West Africa; they represent a disproportionate amount of potential damage to the network when conflict does occur in those regions. The conflict that isolates entire cities is rarer in the North and West compared to the East. Still, when they occur, the damage, quantified in terms of millions of people removed from Africa's integrated economy, tends to be more significant.

{
\begin{figure} 
\centering{
\includegraphics[width=.7\textwidth]{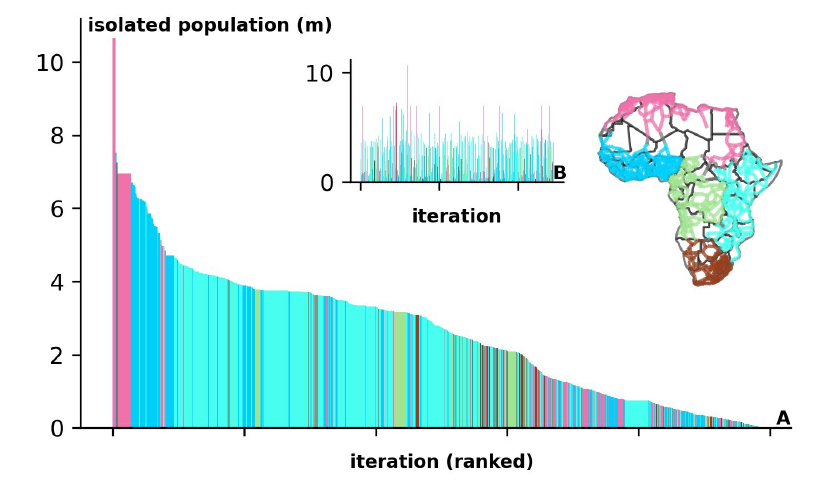}}
\caption{\textbf{A} Results of an attack percolation simulating the effect of limited but cascading conflict, we see that even though West and East Africa are most prone to violence, the combination of a sparse road network and growing conflict on the Horn of Africa results in more isolates — stranded cities with no connections to the broader continent — in the East across simulations. We can interpret thicker bars as repeat events across simulations, showing that much of the largest isolated cities come from rarer events in the North. \textbf{B} The results in order of occurrence during 10.000 runs.}
\label{simulation1}
\end{figure}
}


Results of a cumulative percolation simulating prolonged conflict, cutting edges with probability proportionate to conflict intensity, shows that in some scenarios, it only takes a few hundred events to strand millions of residents, disrupt a quarter or more of interurban traffic and fragment the network (Figure \ref{simulation2}). Although there is little risk of correlated violence across the continent, the threat of it could disrupt trade and discourage integration. As Figures \ref{riskvflow} and \ref{HawkesMap} show, conflict is growing in some areas and persisting in others, leading to prolonged threats that could change the relationships between cities.

{
\begin{figure} 
\centering{
\includegraphics[width=.7\textwidth]{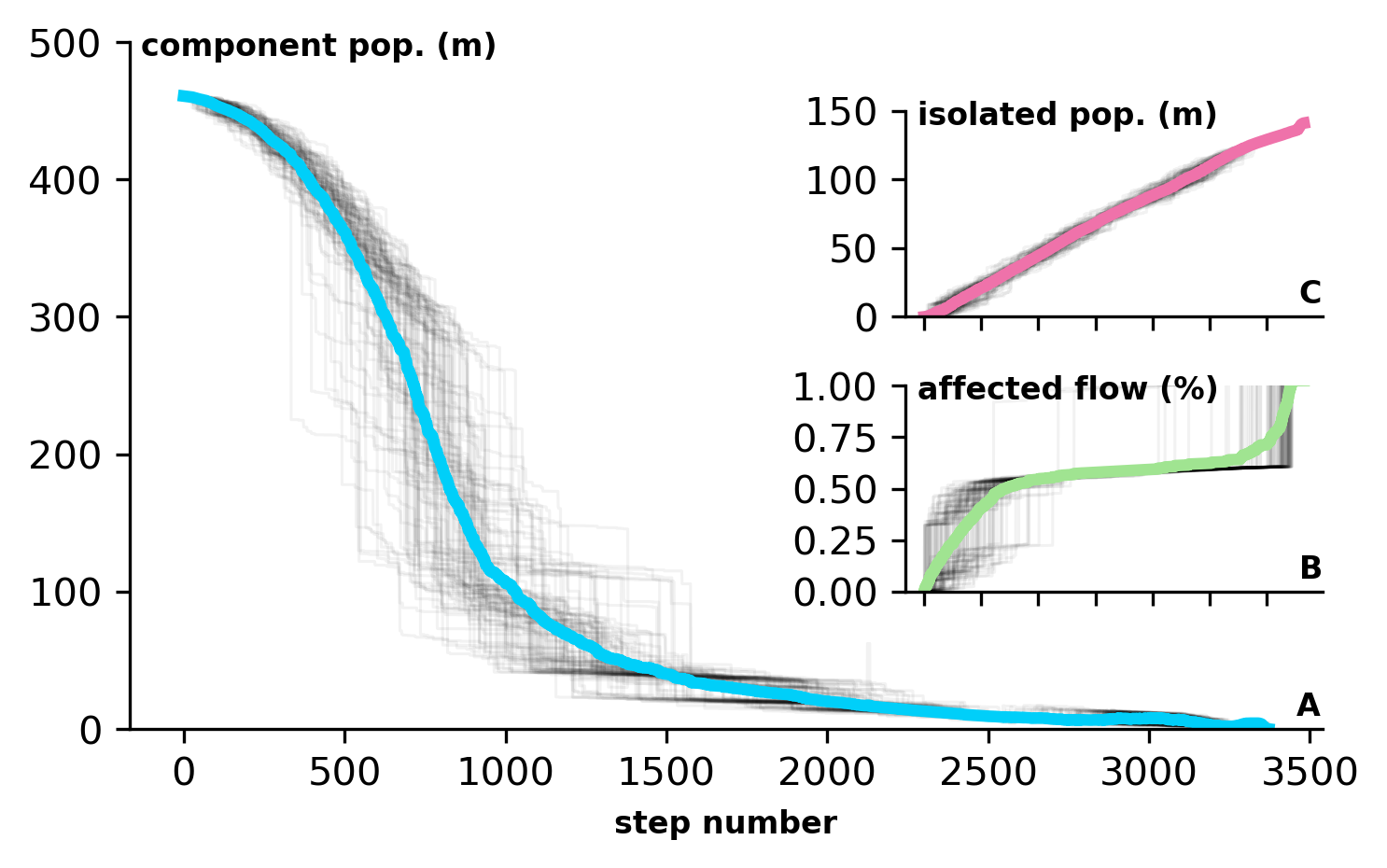}}
\caption{\textbf{A} Number of people in the largest component (vertical axis) as the number of edges is removed (horizontal axis). Each grey line represents a single simulation, while the blue shows the average; the dispersion of grey lines shows that in the worst cases, the component can break during a few hundred conflict events. \textbf{B} The percentage of total flow along the network, according to our gravity model, is disrupted by rising conflict, with some key edges disrupted early on in many of the simulations. \textbf{C} The number of people in cities in isolated nodes at each step, showing monotonic growth across steps with few deviations across simulations, again shown with a grey line.} 
\label{simulation2}
\end{figure}
}

As we move down the risk curve in \ref{simulation2}, we produce maps like those in Figure \ref{percolation}. We remove all links above a given threshold and observe the size of the largest component as we do so, finding that the continent fragments even at high $mu$ values, indicating that we could see effects from violence if it escalates on the continent. Indeed, although the giant connected component is largely unchanged for high $mu$ values, we see isolates at 80 and more at 40. Although the more aggressive scenarios at $mu=20$ and $mu=10$ are less likely without a secular rise in violence across the continent, they are still instructive, as we could see subsets or localised versions of them. 

{
\begin{figure} 
\centering{
\includegraphics[width=.7\textwidth]{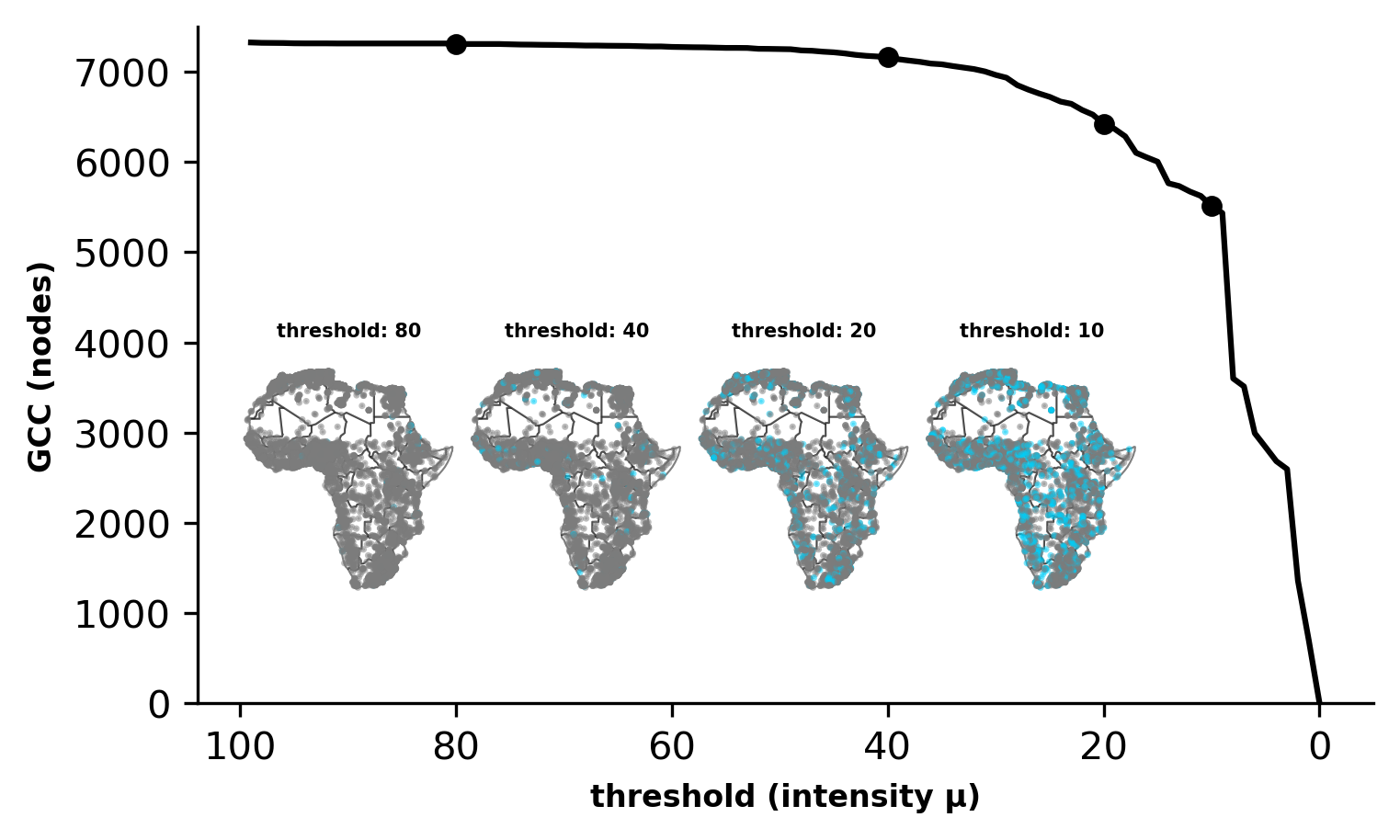}}
\caption{The result cumulative effect of violence on the integrity of Africa's interurban network. Blue here represents new isolates or new components broken off from the largest component. As we remove links at a given level of risk, we see the size of the largest component shrink and the continent fragment into isolated areas.} 
\label{percolation}
\end{figure}
}

\section{Conclusions}

Road transportation is the dominant mode of motorised transport in Africa, accounting for 80\% of goods traffic and 90\% of passenger traffic. However, in 2005, only 22.7\% of the African road network was paved, highlighting the need for infrastructure development in some regions \cite{UNTransportAfrica09}. Even though there is some regional variation in the road density and quality, the road transport network in Africa is sparse and frequently characterised by relatively long delays at borders, high transit tariffs, market access barriers and transport restrictions. These shortcomings are especially prevalent in some of Africa's least developed regions, hindering their economic development efforts \cite{AfricaEUCoop-Transport} and creating a cyclic relationship between poverty and the state of the road network. Furthermore, terrorist groups operating in some regions take advantage of the fragility of the road network to increase their control in an area. Through their attacks, they may seek to create additional disruption in the roads \cite{prieto2020uncovering}, hence exacerbating the already existing issues. It is estimated that between 2001 and 2021, the number of casualties attributed to violence against civilians in Africa increased 260\% \cite{raleigh2010introducing}. 

In this work, we have quantified the impact that disruptions in the road network would have at different locations. The impact measure is based on the magnitude of the traffic flow at a given location, so the impact is higher in locations with more traffic. We have also quantified the risk or intensity at which a location experiences disruption due to violent conflict at a given time. Combining these two measures, we observe that low-risk but high-impact locations are primarily concentrated in North African cities. Importantly, our analysis suggests that high-risk and high-impact cities are concentrated in Nigeria and Ethiopia, leading to higher vulnerability in these countries. A broader band of risk with varying impact stretches across the Sahel. High-risk, low-impact areas are in Somalia and other isolated parts of the network. 

Specifically, we find that due to road disruption caused by violent attacks, a larger population is more likely to become isolated from the rest of the road network in West and East Africa. In addition, if all journeys were to go ahead despite the disruption, the fragmentation of the road network would force a relatively higher percentage of traffic flows to be rerouted in West Africa. This could result in longer travel times, subject to additional delays and congestion and increased vehicle emissions. Further analysis should also focus on the resilience of the road infrastructure by looking at the loss in quality of service, the recovery time and whether the system is improved after any event \cite{bruneau2003framework, roy2019quantifying}.

The network analysis approach that we take here offers several benefits over traditional Geographic Information Systems (GIS) for studying the vulnerability of specific locations within the African road system. The most obvious advantage is the ability of spatial networks to capture the interactions between different places, allowing us to understand the complexity of the system better and ask questions such as `If there is a violent event in a certain location, how does this affect the road transport system as a whole?'. Therefore, our work constitutes a step forward in the direction of understanding the resilience of the road transport network against occurrences of violent conflict in the African continent. Our work also highlights the need to increase the quality and connectivity of the road network in Africa, with the West African region being a priority area.

\section{Supplementary information}

\subsection{The network of all African highways}

{

Primary roads, highways and trunks were obtained from OpenStreetMap  \cite{OpenStreetMap}, and data from Africapolis \cite{Africapolis} was used to obtain the location and size of all urban agglomerations with more than 100,000 inhabitants and also smaller cities near major roads to construct a connected network of all cities. A connected and simplified network of African cities is constructed. Unlike other infrastructure networks (for example, a railway network), the road network is not clearly defined. For example, considering train stations and the railways between them is a natural way to determine the nodes and the edges of the railway network. Road networks need transport nodes representing road crossings.

Cities and road data were spatially aligned and integrated into a single network. Then, disconnected components of the network were joined based on the proximity of the roads \cite{prieto2022constructing, prieto2022detecting}. Then, the network is simplified by dissolving nodes with a degree of two that connect the same type of road at both ends. The node is replaced by a ``longer'' edge, where the length corresponds to the sum of the two edges. The urban network comprises 9,159 edges and 7,361 nodes. The labelled network represents 2,162 cities in the continent and 5,199 transport nodes. The edges correspond to an existing highway between any two nodes. The type of road is used to estimate the speed and travel time to cross it. Any edge with its extremes on different countries is a border crossing, so 120 minutes are added for crossing it. Similarly, extra time for travelling through an urban agglomeration is added, proportional to the population of that city. For each edge, it is identified its road length, whether it crosses an international border and a city in any of its two extremes, resulting in a proxy of the estimated minutes to travel through that edge. 

}

\subsection{The Hawkes process and its parameters}

{
We use a Hawkes point process to model the intensity of events in city $i$. The rate $\mu_i(t)$ for city $i$ at some time $t$ is calculated by
\begin{equation}
\mu_i(t) = \sum_{s:t \geq t_s} \kappa_s \exp{ -\phi (t - t_s)},
\end{equation}
where $t_s$ indicates the time in which event $s$ happened, $\kappa_s$ is the impact of event $s$, and $\phi > 0$ is the model parameter that captures memory decay. With small values of $\phi$, the model keeps the impact for many days, but with high values, the $\mu_i(t)$ has almost no memory of past events. The sum is only for the time $t$ after the event (so $t \geq t_s$), so the values inside the exponential drop to zero as time passes.

We take $\kappa_s$ to be the number of event fatalities. We estimate the values of $\phi$, so the Hawkes process provides the most future information as follows. For some test value of $\phi$, we compute the corresponding Hawkes process for city $i$ and compute the lagged correlation between the $\mu_i(t)$ and the events in city $i$ for the next $t+14$ days. Thus, the objective is that $\mu_i(t)$ should inform as close as possible what will happen in the city two weeks after its value is computed. We compute the correlation between $\mu_i(t)$ and $E_i(t)$, where $E_i(t)$ is the average daily impact experienced in city $i$ between $t$ and $t+14$. A high correlation indicates that $\mu_i(t)$ informs the intensity of future (unknown) events. We add the correlation obtained for all cities into the value $\theta(\phi)$. Then, we vary the parameter $\phi$ and follow the same procedure, obtaining different values of $\theta(\phi)$. We keep the value of $\phi^\star = 0.0987$ since it maximises $\theta^\star(\phi)$ (Figure \ref{PhiTheta}).

\begin{figure} \centering{
\includegraphics[width=0.5\textwidth]{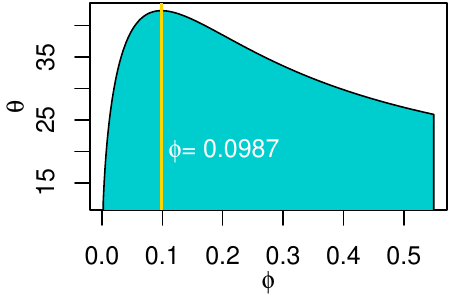}
\caption{Values of $\phi$ (horizontal axis) and its corresponding $\theta(\phi)$ (vertical axis). We keep $\phi^\star = 0.0987$ that maximise the correlation between $\mu_i(t)$ and $E_i(t)$}
 \label{PhiTheta}}
\end{figure}

\begin{figure} \centering{
\includegraphics[width=1\textwidth]{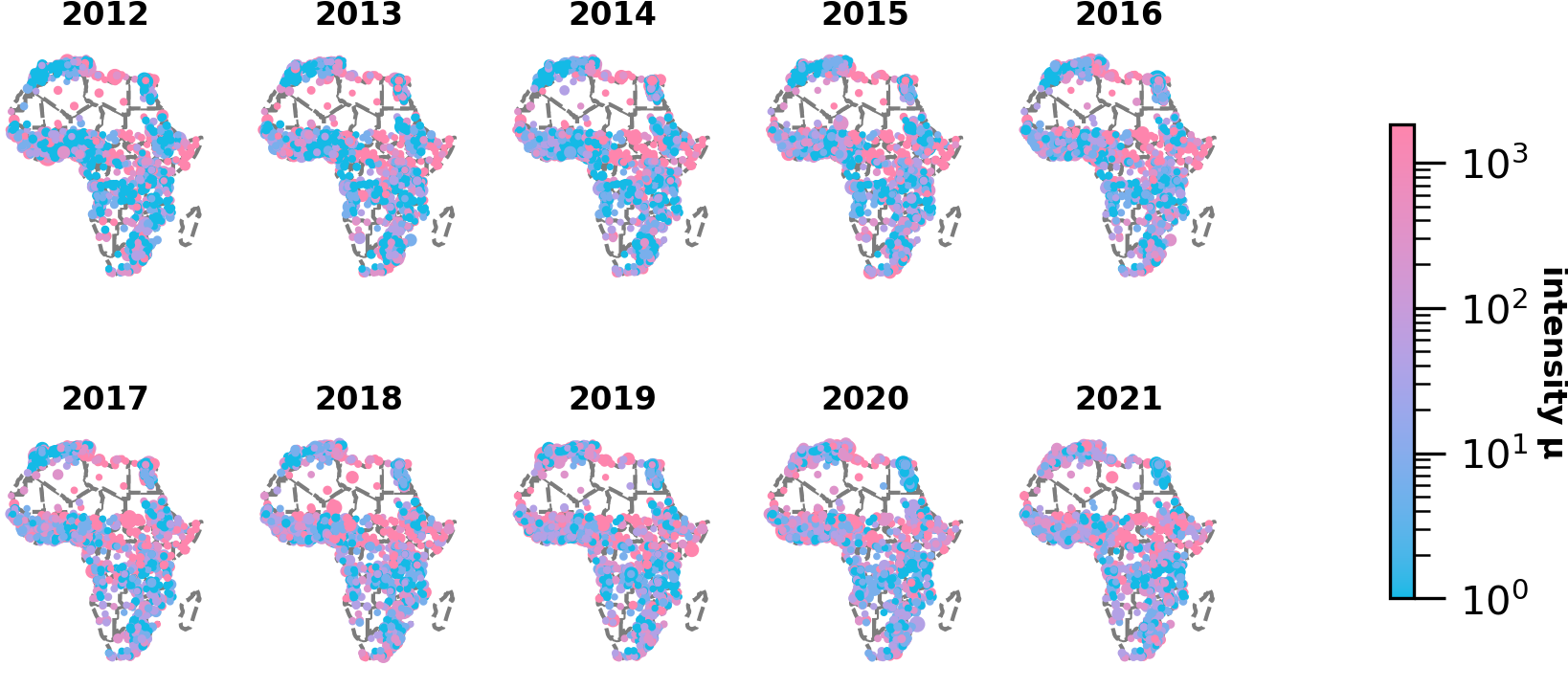}
\caption{Observed values of the intensity of events $\mu_i(t)$ over time across all cities in Africa. } \label{HawkesMap}}
\end{figure}

\begin{figure} \centering{
\includegraphics[width=1\textwidth]{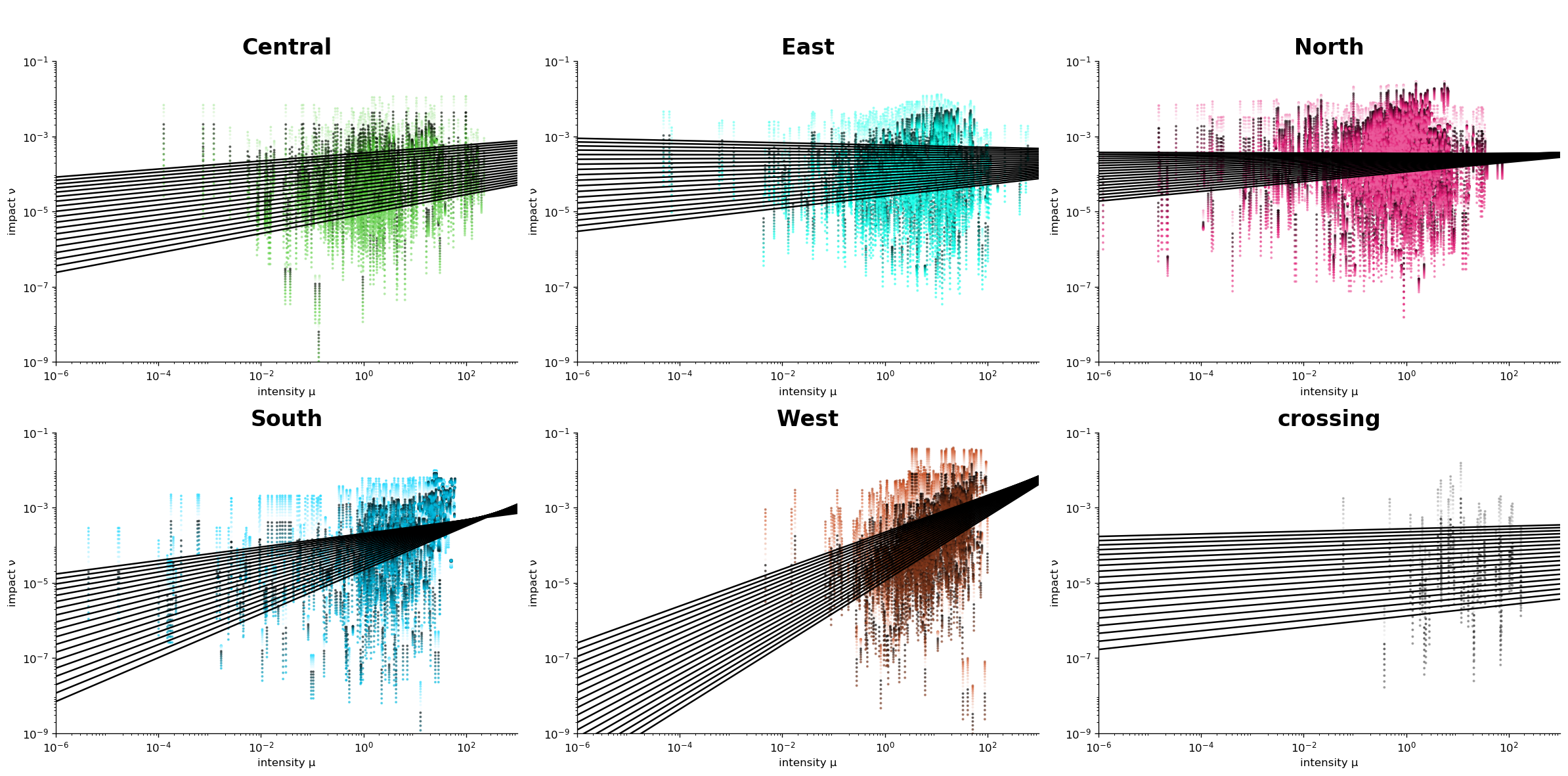}
\caption{We test decay parameters ranging from 1.0 to 4.0 at regular intervals, which correspond to the lowest and highest lines in each plot, respectively. As the effect of distance increases and movement between distant regions decreases, the relationship between flow and conflict disappears in many regions — East and North most notably — but there is still an attenuated relationship in the West, Central and South. The hues here become darker with larger distance parameters.} \label{sensitivity}}
\end{figure}

}

\begin{backmatter}


\section*{Data availability}

The African urban network is available from https://github.com/rafaelprietocuriel/AfricanUrbanNetwork. ACLED data are freely available from https://acleddata.com/data-export-tool/. 

\section*{Competing interests}

The authors declare that they have no competing interests.

\section*{Author's contributions}

AR and RP designed the study and performed the analysis. AR, RP, CC and VM designed the study.  

\section*{Funding}

The research was funded by the Austrian Federal Ministry for Climate Action, Environment, Energy, Mobility, Innovation and Technology (2021-0.664.668) and the Austrian Federal Ministry of the Interior (2022-0.392.231).
 
\bibliography{references} 

\begin{thebibliography}{62}
\ifx \bisbn   \undefined \def \bisbn  #1{ISBN #1}\fi
\ifx \binits  \undefined \def \binits#1{#1}\fi
\ifx \bauthor  \undefined \def \bauthor#1{#1}\fi
\ifx \batitle  \undefined \def \batitle#1{#1}\fi
\ifx \bjtitle  \undefined \def \bjtitle#1{#1}\fi
\ifx \bvolume  \undefined \def \bvolume#1{\textbf{#1}}\fi
\ifx \byear  \undefined \def \byear#1{#1}\fi
\ifx \bissue  \undefined \def \bissue#1{#1}\fi
\ifx \bfpage  \undefined \def \bfpage#1{#1}\fi
\ifx \blpage  \undefined \def \blpage #1{#1}\fi
\ifx \burl  \undefined \def \burl#1{\textsf{#1}}\fi
\ifx \doiurl  \undefined \def \doiurl#1{\textsf{#1}}\fi
\ifx \betal  \undefined \def \betal{\textit{et al.}}\fi
\ifx \binstitute  \undefined \def \binstitute#1{#1}\fi
\ifx \binstitutionaled  \undefined \def \binstitutionaled#1{#1}\fi
\ifx \bctitle  \undefined \def \bctitle#1{#1}\fi
\ifx \beditor  \undefined \def \beditor#1{#1}\fi
\ifx \bpublisher  \undefined \def \bpublisher#1{#1}\fi
\ifx \bbtitle  \undefined \def \bbtitle#1{#1}\fi
\ifx \bedition  \undefined \def \bedition#1{#1}\fi
\ifx \bseriesno  \undefined \def \bseriesno#1{#1}\fi
\ifx \blocation  \undefined \def \blocation#1{#1}\fi
\ifx \bsertitle  \undefined \def \bsertitle#1{#1}\fi
\ifx \bsnm \undefined \def \bsnm#1{#1}\fi
\ifx \bsuffix \undefined \def \bsuffix#1{#1}\fi
\ifx \bparticle \undefined \def \bparticle#1{#1}\fi
\ifx \barticle \undefined \def \barticle#1{#1}\fi
\ifx \bconfdate \undefined \def \bconfdate #1{#1}\fi
\ifx \botherref \undefined \def \botherref #1{#1}\fi
\ifx \url \undefined \def \url#1{\textsf{#1}}\fi
\ifx \bchapter \undefined \def \bchapter#1{#1}\fi
\ifx \bbook \undefined \def \bbook#1{#1}\fi
\ifx \bcomment \undefined \def \bcomment#1{#1}\fi
\ifx \oauthor \undefined \def \oauthor#1{#1}\fi
\ifx \citeauthoryear \undefined \def \citeauthoryear#1{#1}\fi
\ifx \endbibitem  \undefined \def \endbibitem {}\fi
\ifx \bconflocation  \undefined \def \bconflocation#1{#1}\fi
\ifx \arxivurl  \undefined \def \arxivurl#1{\textsf{#1}}\fi
\csname PreBibitemsHook\endcsname

\bibitem{ganin2017resilience}
\begin{barticle}
\bauthor{\bsnm{Ganin}, \binits{A.A.}},
\bauthor{\bsnm{Kitsak}, \binits{M.}},
\bauthor{\bsnm{Marchese}, \binits{D.}},
\bauthor{\bsnm{Keisler}, \binits{J.M.}},
\bauthor{\bsnm{Seager}, \binits{T.}},
\bauthor{\bsnm{Linkov}, \binits{I.}}:
\batitle{Resilience and efficiency in transportation networks}.
\bjtitle{Science Advances}
\bvolume{3}(\bissue{12}),
\bfpage{1701079}
(\byear{2017})
\end{barticle}
\endbibitem

\bibitem{maureira2017everyday}
\begin{bchapter}
\bauthor{\bsnm{{Marin-Maureira}}, \binits{V.}},
\bauthor{\bsnm{{Karimi}}, \binits{K.}}:
\bctitle{The everyday and the post-disaster urban systems as one thing: A configurational approach to enhance the recovery and resilience of cities affected by tsunamis.}
In: \bbtitle{Proceedings-11th International Space Syntax Symposium, SSS 2017},
vol. \bseriesno{11},
pp. \bfpage{91}--\blpage{1}
(\byear{2017}).
\bcomment{Instituto Superior T{\'e}cnico, Portugal}
\end{bchapter}
\endbibitem

\bibitem{roy2019quantifying}
\begin{barticle}
\bauthor{\bsnm{{Roy}}, \binits{K.C.}},
\bauthor{\bsnm{{Cebrian}}, \binits{M.}},
\bauthor{\bsnm{{Hasan}}, \binits{S.}}:
\batitle{Quantifying human mobility resilience to extreme events using geo-located social media data}.
\bjtitle{EPJ Data Science}
\bvolume{8}(\bissue{1}),
\bfpage{1}--\blpage{15}
(\byear{2019})
\end{barticle}
\endbibitem

\bibitem{raleigh2010introducing}
\begin{barticle}
\bauthor{\bsnm{{Raleigh}}, \binits{C.}},
\bauthor{\bsnm{{Linke}}, \binits{A.}},
\bauthor{\bsnm{{Hegre}}, \binits{H.}},
\bauthor{\bsnm{{Karlsen}}, \binits{J.}}:
\batitle{Introducing {ACLED}: an {A}rmed {C}onflict {L}ocation and {E}vent {D}ataset: special data feature}.
\bjtitle{Journal of Peace Research}
\bvolume{47}(\bissue{5}),
\bfpage{651}--\blpage{660}
(\byear{2010})
\end{barticle}
\endbibitem

\bibitem{chathamTerrorism}
\begin{botherref}
\oauthor{\bsnm{Vines}, \binits{A.}},
\oauthor{\bsnm{Wallace}, \binits{J.}}:
Terrorism in africa.
Technical report,
Chatham House
(2021).
\url{https://www.chathamhouse.org/2021/09/terrorism-africa}
\end{botherref}
\endbibitem

\bibitem{burke2015climate}
\begin{barticle}
\bauthor{\bsnm{Burke}, \binits{M.}},
\bauthor{\bsnm{Hsiang}, \binits{S.M.}},
\bauthor{\bsnm{Miguel}, \binits{E.}}:
\batitle{Climate and conflict}.
\bjtitle{Annu. Rev. Econ.}
\bvolume{7}(\bissue{1}),
\bfpage{577}--\blpage{617}
(\byear{2015})
\end{barticle}
\endbibitem

\bibitem{harari2018conflict}
\begin{barticle}
\bauthor{\bsnm{Harari}, \binits{M.}},
\bauthor{\bsnm{Ferrara}, \binits{E.L.}}:
\batitle{Conflict, climate, and cells: a disaggregated analysis}.
\bjtitle{Review of Economics and Statistics}
\bvolume{100}(\bissue{4}),
\bfpage{594}--\blpage{608}
(\byear{2018})
\end{barticle}
\endbibitem

\bibitem{Africapolis}
\begin{botherref}
\oauthor{\bsnm{OECD/SWAC}}:
Africapolis (database).
Organisation for Economic Co-operation and Development.
Accessed: September 2019
(2018)
\end{botherref}
\endbibitem

\bibitem{OpenStreetMap}
\begin{botherref}
\oauthor{\bsnm{{OpenStreetMap contributors}}}:
{Planet dump retrieved from https://planet.osm.org }.
\url{ https://www.openstreetmap.org }
(2021)
\end{botherref}
\endbibitem

\bibitem{prieto2022detecting}
\begin{barticle}
\bauthor{\bsnm{{Prieto-Curiel}}, \binits{R.}},
\bauthor{\bsnm{{Schumann}}, \binits{A.}},
\bauthor{\bsnm{{Heo}}, \binits{I.}},
\bauthor{\bsnm{{Heinrigs}}, \binits{P.}}:
\batitle{Detecting cities with high intermediacy in the {A}frican urban network}.
\bjtitle{Computers, Environment and Urban Systems}
\bvolume{98},
\bfpage{101869}
(\byear{2022})
\end{barticle}
\endbibitem

\bibitem{prieto2022constructing}
\begin{barticle}
\bauthor{\bsnm{{Prieto-Curiel}}, \binits{R.}},
\bauthor{\bsnm{{Heo}}, \binits{I.}},
\bauthor{\bsnm{{Schumann}}, \binits{A.}},
\bauthor{\bsnm{{Heinrigs}}, \binits{P.}}:
\batitle{Constructing a simplified interurban road network based on crowdsourced geodata}.
\bjtitle{MethodsX}
\bvolume{9},
\bfpage{101845}
(\byear{2022})
\end{barticle}
\endbibitem

\bibitem{ZipfTerrorism}
\begin{barticle}
\bauthor{\bsnm{{Clauset}}, \binits{A.}},
\bauthor{\bsnm{{Young}}, \binits{M.}},
\bauthor{\bsnm{{Gleditsch}}, \binits{K.S.}}:
\batitle{On the frequency of severe terrorist events}.
\bjtitle{Journal of Conflict Resolution}
\bvolume{51}(\bissue{1}),
\bfpage{58}--\blpage{87}
(\byear{2007})
\end{barticle}
\endbibitem

\bibitem{guo2019common}
\begin{botherref}
\oauthor{\bsnm{{Guo}}, \binits{W.}}:
Common statistical patterns in urban terrorism.
Royal Society Open Science
\textbf{6}
(2019)
\end{botherref}
\endbibitem

\bibitem{radil2022urban}
\begin{botherref}
\oauthor{\bsnm{{Radil}}, \binits{S.}},
\oauthor{\bsnm{{Walther}}, \binits{O.}},
\oauthor{\bsnm{{Dorward}}, \binits{N.}},
\oauthor{\bsnm{{Pflaum}}, \binits{M.}}:
Urban-rural geographies of political violence in {N}orth and {W}est {A}frica.
Available at SSRN
(2022)
\end{botherref}
\endbibitem

\bibitem{ViolenceNorthWestAfrica}
\begin{barticle}
\bauthor{\bsnm{{Skillicorn}}, \binits{D.B.}},
\bauthor{\bsnm{{Walther}}, \binits{O.}},
\bauthor{\bsnm{{Leuprecht}}, \binits{C.}},
\bauthor{\bsnm{{Zheng}}, \binits{Q.}}:
\batitle{The diffusion and permeability of political violence in {N}orth and {W}est {A}frica}.
\bjtitle{Terrorism and Political Violence}
\bvolume{0}(\bissue{0}),
\bfpage{1}--\blpage{23}
(\byear{2019}).
doi:\doiurl{10.1080/09546553.2019.1598388}.
\arxivurl{https://doi.org/10.1080/09546553.2019.1598388}
\end{barticle}
\endbibitem

\bibitem{buhaug2008contagion}
\begin{barticle}
\bauthor{\bsnm{{Buhaug}}, \binits{H.}},
\bauthor{\bsnm{{Gleditsch}}, \binits{K.S.}}:
\batitle{Contagion or confusion? why conflicts cluster in space}.
\bjtitle{International Studies Quarterly}
\bvolume{52}(\bissue{2}),
\bfpage{215}--\blpage{233}
(\byear{2008})
\end{barticle}
\endbibitem

\bibitem{LawCrimeConcentration}
\begin{barticle}
\bauthor{\bsnm{{Weisburd}}, \binits{D.}}:
\batitle{The law of crime concentration and the criminology of place}.
\bjtitle{Criminology}
\bvolume{53}(\bissue{2}),
\bfpage{133}--\blpage{157}
(\byear{2015}).
doi:\doiurl{10.1111/1745-9125.12070}
\end{barticle}
\endbibitem

\bibitem{CrimeConcentrationVaryingCitySize}
\begin{botherref}
\oauthor{\bsnm{{Hipp}}, \binits{J.R.}},
\oauthor{\bsnm{{Kim}}, \binits{Y.-A.}}:
Measuring crime concentration across cities of varying sizes: Complications based on the spatial and temporal scale employed.
Journal of Quantitative Criminology,
1--38
(2016).
doi:\doiurl{10.1007/s10940-016-9328-3}
\end{botherref}
\endbibitem

\bibitem{hawkes1971spectra}
\begin{barticle}
\bauthor{\bsnm{{Hawkes}}, \binits{A.G.}}:
\batitle{Spectra of some self-exciting and mutually exciting point processes}.
\bjtitle{Biometrika}
\bvolume{58}(\bissue{1}),
\bfpage{83}--\blpage{90}
(\byear{1971})
\end{barticle}
\endbibitem

\bibitem{chuang2019mathematical}
\begin{botherref}
\oauthor{\bsnm{{Chuang}}, \binits{Y.-l.}},
\oauthor{\bsnm{{D'Orsogna}}, \binits{M.R.}}:
Mathematical models of radicalization and terrorism.
ArXiv preprint arXiv:1903.08485
(2019)
\end{botherref}
\endbibitem

\bibitem{MohlerExcitingPointProcess}
\begin{botherref}
\oauthor{\bsnm{{Mohler}}, \binits{G.O.}},
\oauthor{\bsnm{{Short}}, \binits{M.B.}},
\oauthor{\bsnm{{Brantingham}}, \binits{P.J.}},
\oauthor{\bsnm{{Schoenberg}}, \binits{F.P.}},
\oauthor{\bsnm{{Tita}}, \binits{G.E.}}:
Self-exciting point process modeling of crime.
Journal of the American Statistical Association
(2012)
\end{botherref}
\endbibitem

\bibitem{laub2015hawkes}
\begin{botherref}
\oauthor{\bsnm{{Laub}}, \binits{P.J.}},
\oauthor{\bsnm{{Taimre}}, \binits{T.}},
\oauthor{\bsnm{{Pollett}}, \binits{P.K.}}:
Hawkes processes.
arXiv preprint arXiv:1507.02822
(2015)
\end{botherref}
\endbibitem

\bibitem{chowdhary2023temporal}
\begin{barticle}
\bauthor{\bsnm{{Chowdhary}}, \binits{S.}},
\bauthor{\bsnm{{Andres}}, \binits{E.}},
\bauthor{\bsnm{{Manna}}, \binits{A.}},
\bauthor{\bsnm{{Blagojevi{\'c}}}, \binits{L.}},
\bauthor{\bsnm{{Di Gaetano}}, \binits{L.}},
\bauthor{\bsnm{{I{\~n}iguez}}, \binits{G.}}:
\batitle{Temporal patterns of reciprocity in communication networks}.
\bjtitle{EPJ Data Science}
\bvolume{12}(\bissue{1}),
\bfpage{7}
(\byear{2023})
\end{barticle}
\endbibitem

\bibitem{porter2012self}
\begin{botherref}
\oauthor{\bsnm{{Porter}}, \binits{M.D.}},
\oauthor{\bsnm{{White}}, \binits{G.}}:
Self-exciting hurdle models for terrorist activity.
The Annals of Applied Statistics,
106--124
(2012)
\end{botherref}
\endbibitem

\bibitem{yuan2019fast}
\begin{botherref}
\oauthor{\bsnm{{Yuan}}, \binits{B.}},
\oauthor{\bsnm{{Schoenberg}}, \binits{F.P.}},
\oauthor{\bsnm{{Bertozzi}}, \binits{A.L.}}:
Fast estimation of multivariate spatiotemporal {H}awkes processes and network reconstruction.
.
(2019)
\end{botherref}
\endbibitem

\bibitem{telesca2006global}
\begin{barticle}
\bauthor{\bsnm{{Telesca}}, \binits{L.}},
\bauthor{\bsnm{{Lovallo}}, \binits{M.}}:
\batitle{Are global terrorist attacks time-correlated?}
\bjtitle{Physica A: Statistical Mechanics and its Applications}
\bvolume{362}(\bissue{2}),
\bfpage{480}--\blpage{484}
(\byear{2006})
\end{barticle}
\endbibitem

\bibitem{clauset2013estimating}
\begin{barticle}
\bauthor{\bsnm{{Clauset}}, \binits{A.}},
\bauthor{\bsnm{{Woodard}}, \binits{R.}}, \betal:
\batitle{Estimating the historical and future probabilities of large terrorist events}.
\bjtitle{Annals of Applied Statistics}
\bvolume{7}(\bissue{4}),
\bfpage{1838}--\blpage{1865}
(\byear{2013})
\end{barticle}
\endbibitem

\bibitem{crucitti2004error}
\begin{barticle}
\bauthor{\bsnm{Crucitti}, \binits{P.}},
\bauthor{\bsnm{Latora}, \binits{V.}},
\bauthor{\bsnm{Marchiori}, \binits{M.}},
\bauthor{\bsnm{Rapisarda}, \binits{A.}}:
\batitle{Error and attack tolerance of complex networks}.
\bjtitle{Physica A: Statistical mechanics and its applications}
\bvolume{340}(\bissue{1-3}),
\bfpage{388}--\blpage{394}
(\byear{2004})
\end{barticle}
\endbibitem

\bibitem{cascadingPorta2004}
\begin{barticle}
\bauthor{\bsnm{Crucitti}, \binits{P.}},
\bauthor{\bsnm{Latora}, \binits{V.}},
\bauthor{\bsnm{Marchiori}, \binits{M.}}:
\batitle{Model for cascading failures in complex networks}.
\bjtitle{Phys. Rev. E}
\bvolume{69},
\bfpage{045104}
(\byear{2004}).
doi:\doiurl{10.1103/PhysRevE.69.045104}
\end{barticle}
\endbibitem

\bibitem{newman2003structure}
\begin{barticle}
\bauthor{\bsnm{Newman}, \binits{M.E.}}:
\batitle{The structure and function of complex networks}.
\bjtitle{SIAM review}
\bvolume{45}(\bissue{2}),
\bfpage{167}--\blpage{256}
(\byear{2003})
\end{barticle}
\endbibitem

\bibitem{barabasi2016network}
\begin{bbook}
\bauthor{\bsnm{Barab{\'a}si}, \binits{A.-L.}}, \betal:
\bbtitle{Network Science}.
\bpublisher{Cambridge University Press},
\blocation{Cambridge, UK}
(\byear{2016})
\end{bbook}
\endbibitem

\bibitem{Gallos2011}
\begin{botherref}
\oauthor{\bsnm{Gallos}, \binits{L.K.}},
\oauthor{\bsnm{Makse}, \binits{H.A.}},
\oauthor{\bsnm{Sigman}, \binits{M.}}:
{A small-world of weak ties provides optimal global integration of self-similar modules in functional brain networks}
\textbf{109}(8),
2825--2830
(2011).
doi:\doiurl{10.1073/pnas.1106612109}
\end{botherref}
\endbibitem

\bibitem{newman2002spread}
\begin{barticle}
\bauthor{\bsnm{Newman}, \binits{M.E.}}:
\batitle{Spread of epidemic disease on networks}.
\bjtitle{Physical Review E}
\bvolume{66}(\bissue{1}),
\bfpage{016128}
(\byear{2002})
\end{barticle}
\endbibitem

\bibitem{Gallos2012}
\begin{barticle}
\bauthor{\bsnm{Gallos}, \binits{L.K.}},
\bauthor{\bsnm{Barttfeld}, \binits{P.}},
\bauthor{\bsnm{Havlin}, \binits{S.}},
\bauthor{\bsnm{Sigman}, \binits{M.}},
\bauthor{\bsnm{Makse}, \binits{H.A.}}:
\batitle{Collective behavior in the spatial spreading of obesity}.
\bjtitle{Scientific Reports}
\bvolume{2}(\bissue{1}),
\bfpage{1}--\blpage{9}
(\byear{2012})
\end{barticle}
\endbibitem

\bibitem{ArcautePercolation16}
\begin{barticle}
\bauthor{\bsnm{{Arcaute}}, \binits{E.}},
\bauthor{\bsnm{{Molinero}}, \binits{C.}},
\bauthor{\bsnm{{Hatna}}, \binits{E.}},
\bauthor{\bsnm{{Murcio}}, \binits{R.}},
\bauthor{\bsnm{{Vargas-Ruiz}}, \binits{C.}},
\bauthor{\bsnm{{Masucci}}, \binits{A.P.}},
\bauthor{\bsnm{{Batty}}, \binits{M.}}:
\batitle{Cities and regions in {B}ritain through hierarchical percolation}.
\bjtitle{Royal Society Open Science}
\bvolume{3}(\bissue{4}),
\bfpage{150691}
(\byear{2016}).
doi:\doiurl{10.1098/rsos.150691}.
\arxivurl{https://royalsocietypublishing.org/doi/pdf/10.1098/rsos.150691}
\end{barticle}
\endbibitem

\bibitem{Fluschnik2016}
\begin{barticle}
\bauthor{\bsnm{Fluschnik}, \binits{T.}},
\bauthor{\bsnm{Kriewald}, \binits{S.}},
\bauthor{\bsnm{Ros}, \binits{A.G.C.}},
\bauthor{\bsnm{Zhou}, \binits{B.}},
\bauthor{\bsnm{Reusser}, \binits{D.E.}},
\bauthor{\bsnm{Kropp}, \binits{J.P.}},
\bauthor{\bsnm{Rybski}, \binits{D.}}:
\batitle{{The size distribution, scaling properties and spatial organization of urban clusters: A global and regional percolation perspective}}.
\bjtitle{ISPRS International Journal of Geo-Information}
\bvolume{5}(\bissue{7}),
\bfpage{1}--\blpage{14}
(\byear{2016}).
doi:\doiurl{10.3390/ijgi5070110}.
\arxivurl{1404.0353}
\end{barticle}
\endbibitem

\bibitem{Piovani2017}
\begin{botherref}
\oauthor{\bsnm{Piovani}, \binits{D.}},
\oauthor{\bsnm{Molinero}, \binits{C.}},
\oauthor{\bsnm{Wilson}, \binits{A.}}:
{Urban retail location: Insights from percolation theory and spatial interaction modeling}.
PLoS ONE
\textbf{12}(10)
(2017).
doi:\doiurl{10.1371/journal.pone.0185787}.
\arxivurl{1703.10419}
\end{botherref}
\endbibitem

\bibitem{Behnisch2019}
\begin{botherref}
\oauthor{\bsnm{Behnisch}, \binits{M.}},
\oauthor{\bsnm{Schorcht}, \binits{M.}},
\oauthor{\bsnm{Kriewald}, \binits{S.}},
\oauthor{\bsnm{Rybski}, \binits{D.}}:
{Settlement percolation: A study of building connectivity and poles of inaccessibility}.
Landscape and Urban Planning
\textbf{191}(June)
(2019).
doi:\doiurl{10.1016/j.landurbplan.2019.103631}
\end{botherref}
\endbibitem

\bibitem{marin2023scalar}
\begin{botherref}
\oauthor{\bsnm{Marin}, \binits{V.}},
\oauthor{\bsnm{Molinero}, \binits{C.}},
\oauthor{\bsnm{Arcaute}, \binits{E.}}:
The scalar mismatch of regional governance: a comparative analysis of hierarchical structures
(2023).
\arxivurl{2302.04904}
\end{botherref}
\endbibitem

\bibitem{Callaway2000}
\begin{barticle}
\bauthor{\bsnm{Callaway}, \binits{D.S.}},
\bauthor{\bsnm{Newman}, \binits{M.E.J.}},
\bauthor{\bsnm{Strogatz}, \binits{S.H.}},
\bauthor{\bsnm{Watts}, \binits{D.J.}}:
\batitle{Network robustness and fragility: Percolation on random graphs}.
\bjtitle{Phys. Rev. Lett.}
\bvolume{85},
\bfpage{5468}--\blpage{5471}
(\byear{2000}).
doi:\doiurl{10.1103/PhysRevLett.85.5468}
\end{barticle}
\endbibitem

\bibitem{gil2008flood}
\begin{barticle}
\bauthor{\bsnm{Gil}, \binits{J.}},
\bauthor{\bsnm{Steinbach}, \binits{P.}}:
\batitle{From flood risk to indirect flood impact: evaluation of street network performance for effective management, response and repair}.
\bjtitle{WIT Transactions on Ecology and the Environment}
\bvolume{118},
\bfpage{335}--\blpage{344}
(\byear{2008})
\end{barticle}
\endbibitem

\bibitem{PaoloMasucci2016}
\begin{barticle}
\bauthor{\bsnm{Masucci}, \binits{P.}},
\bauthor{\bsnm{Molinero}, \binits{C.}}:
\batitle{Robustness and closeness centrality for self-organized and planned cities}.
\bjtitle{The European Physical Journal B}
\bvolume{89}(\bissue{2}),
\bfpage{53}
(\byear{2016}).
doi:\doiurl{10.1140/epjb/e2016-60431-2}
\end{barticle}
\endbibitem

\bibitem{berche2009resilience}
\begin{barticle}
\bauthor{\bsnm{Berche}, \binits{B.}},
\bauthor{\bparticle{von} \bsnm{Ferber}, \binits{C.}},
\bauthor{\bsnm{Holovatch}, \binits{T.}},
\bauthor{\bsnm{Holovatch}, \binits{Y.}}:
\batitle{Resilience of public transport networks against attacks}.
\bjtitle{The European Physical Journal B}
\bvolume{71}(\bissue{1}),
\bfpage{125}--\blpage{137}
(\byear{2009})
\end{barticle}
\endbibitem

\bibitem{Derrible2010}
\begin{barticle}
\bauthor{\bsnm{Derrible}, \binits{S.}},
\bauthor{\bsnm{Kennedy}, \binits{C.}}:
\batitle{The complexity and robustness of metro networks}.
\bjtitle{Physica A: Statistical Mechanics and its Applications}
\bvolume{389}(\bissue{17}),
\bfpage{3678}--\blpage{3691}
(\byear{2010}).
doi:\doiurl{10.1016/j.physa.2010.04.008}
\end{barticle}
\endbibitem

\bibitem{pagani2019resilience}
\begin{barticle}
\bauthor{\bsnm{Pagani}, \binits{A.}},
\bauthor{\bsnm{Mosquera}, \binits{G.}},
\bauthor{\bsnm{Alturki}, \binits{A.}},
\bauthor{\bsnm{Johnson}, \binits{S.}},
\bauthor{\bsnm{Jarvis}, \binits{S.}},
\bauthor{\bsnm{Wilson}, \binits{A.}},
\bauthor{\bsnm{Guo}, \binits{W.}},
\bauthor{\bsnm{Varga}, \binits{L.}}:
\batitle{Resilience or robustness: identifying topological vulnerabilities in rail networks}.
\bjtitle{Royal Society Open Science}
\bvolume{6}(\bissue{2}),
\bfpage{181301}
(\byear{2019})
\end{barticle}
\endbibitem

\bibitem{Holme2002}
\begin{barticle}
\bauthor{\bsnm{Holme}, \binits{P.}},
\bauthor{\bsnm{Kim}, \binits{B.J.}},
\bauthor{\bsnm{Yoon}, \binits{C.N.}},
\bauthor{\bsnm{Han}, \binits{S.K.}}:
\batitle{Attack vulnerability of complex networks}.
\bjtitle{Phys. Rev. E}
\bvolume{65},
\bfpage{056109}
(\byear{2002}).
doi:\doiurl{10.1103/PhysRevE.65.056109}
\end{barticle}
\endbibitem

\bibitem{meijers2016borrowing}
\begin{barticle}
\bauthor{\bsnm{{Meijers}}, \binits{E.J.}},
\bauthor{\bsnm{{Burger}}, \binits{M.J.}},
\bauthor{\bsnm{{Hoogerbrugge}}, \binits{M.M.}}:
\batitle{Borrowing size in networks of cities: City size, network connectivity and metropolitan functions in {E}urope}.
\bjtitle{Papers in Regional Science}
\bvolume{95}(\bissue{1}),
\bfpage{181}--\blpage{198}
(\byear{2016})
\end{barticle}
\endbibitem

\bibitem{StreetNetw}
\begin{barticle}
\bauthor{\bsnm{{Ozuduru}}, \binits{B.H.}},
\bauthor{\bsnm{{Webster}}, \binits{C.J.}},
\bauthor{\bsnm{{Chiaradia}}, \binits{A.J.F.}},
\bauthor{\bsnm{{Yucesoy}}, \binits{E.}}:
\batitle{Associating street-network centrality with spontaneous and planned subcentres}.
\bjtitle{Urban Studies}
\bvolume{58}(\bissue{10}),
\bfpage{2059}--\blpage{2078}
(\byear{2021}).
doi:\doiurl{10.1177/0042098020931302}
\end{barticle}
\endbibitem

\bibitem{barbosa2018human}
\begin{barticle}
\bauthor{\bsnm{{Barbosa}}, \binits{H.}},
\bauthor{\bsnm{{Barthelemy}}, \binits{M.}},
\bauthor{\bsnm{{Ghoshal}}, \binits{G.}},
\bauthor{\bsnm{{James}}, \binits{C.R.}},
\bauthor{\bsnm{{Lenormand}}, \binits{M.}},
\bauthor{\bsnm{{Louail}}, \binits{T.}},
\bauthor{\bsnm{{Menezes}}, \binits{R.}},
\bauthor{\bsnm{{Ramasco}}, \binits{J.J.}},
\bauthor{\bsnm{{Simini}}, \binits{F.}},
\bauthor{\bsnm{{Tomasini}}, \binits{M.}}:
\batitle{Human mobility: Models and applications}.
\bjtitle{Physics Reports}
\bvolume{734},
\bfpage{1}--\blpage{74}
(\byear{2018})
\end{barticle}
\endbibitem

\bibitem{buechel2013dynamics}
\begin{barticle}
\bauthor{\bsnm{{Buechel}}, \binits{B.}},
\bauthor{\bsnm{{Buskens}}, \binits{V.}}:
\batitle{The dynamics of closeness and betweenness}.
\bjtitle{The Journal of Mathematical Sociology}
\bvolume{37}(\bissue{3}),
\bfpage{159}--\blpage{191}
(\byear{2013})
\end{barticle}
\endbibitem

\bibitem{singh2020node}
\begin{barticle}
\bauthor{\bsnm{{Singh}}, \binits{A.}},
\bauthor{\bsnm{{Singh}}, \binits{R.R.}},
\bauthor{\bsnm{{Iyengar}}, \binits{S.}}:
\batitle{Node-weighted centrality: a new way of centrality hybridization}.
\bjtitle{Computational Social Networks}
\bvolume{7}(\bissue{1}),
\bfpage{1}--\blpage{33}
(\byear{2020})
\end{barticle}
\endbibitem

\bibitem{freeman1978centrality}
\begin{barticle}
\bauthor{\bsnm{{Freeman}}, \binits{L.C.}}:
\batitle{Centrality in social networks conceptual clarification}.
\bjtitle{Social Networks}
\bvolume{1}(\bissue{3}),
\bfpage{215}--\blpage{239}
(\byear{1978})
\end{barticle}
\endbibitem

\bibitem{Gravity}
\begin{botherref}
\oauthor{\bsnm{{Anderson}}, \binits{J.E.}}:
The gravity model.
Technical report,
National Bureau of Economic Research
(2010)
\end{botherref}
\endbibitem

\bibitem{baggag2018resilience}
\begin{barticle}
\bauthor{\bsnm{{Baggag}}, \binits{A.}},
\bauthor{\bsnm{{Abbar}}, \binits{S.}},
\bauthor{\bsnm{{Zanouda}}, \binits{T.}},
\bauthor{\bsnm{{Srivastava}}, \binits{J.}}:
\batitle{Resilience analytics: coverage and robustness in multi-modal transportation networks}.
\bjtitle{EPJ Data Science}
\bvolume{7},
\bfpage{1}--\blpage{21}
(\byear{2018})
\end{barticle}
\endbibitem

\bibitem{schuster2023stress}
\begin{botherref}
\oauthor{\bsnm{{Schuster}}, \binits{H.}},
\oauthor{\bsnm{{Polleres}}, \binits{A.}},
\oauthor{\bsnm{{Wachs}}, \binits{J.}}:
Stress-testing road networks and access to medical care.
arXiv preprint arXiv:2307.02250
(2023)
\end{botherref}
\endbibitem

\bibitem{/content/publication/9789264265875-en}
\begin{bbook}
\bauthor{\bsnm{OECD}},
\bauthor{\bsnm{Sahel}},
\bauthor{\bsnm{Club}, \binits{W.A.}}:
\bbtitle{Cross-border Co-operation and Policy Networks in West Africa},
p. \bfpage{224}
(\byear{2017}).
doi:\doiurl{10.1787/9789264265875-en}.
\burl{https://www.oecd-ilibrary.org/content/publication/9789264265875-en}
\end{bbook}
\endbibitem

\bibitem{mantel1967detection}
\begin{barticle}
\bauthor{\bsnm{Mantel}, \binits{N.}}:
\batitle{The detection of disease clustering and a generalized regression approach}.
\bjtitle{Cancer research}
\bvolume{27}(\bissue{2\_Part\_1}),
\bfpage{209}--\blpage{220}
(\byear{1967})
\end{barticle}
\endbibitem

\bibitem{Knox}
\begin{barticle}
\bauthor{\bsnm{Knox}, \binits{E.G.}}:
\batitle{{The Detection of Space‐Time Interactions}}.
\bjtitle{Journal of the Royal Statistical Society Series C}
\bvolume{13}(\bissue{1}),
\bfpage{25}--\blpage{29}
(\byear{1964}).
doi:\doiurl{10.2307/2985220}
\end{barticle}
\endbibitem

\bibitem{UNTransportAfrica09}
\begin{botherref}
\oauthor{\bsnm{Economic}, \binits{U.N.-}},
\oauthor{\bparticle{for} \bsnm{Africa}, \binits{S.C.-E.C.}}:
Africa review report on transport.
Technical report,
United Nations,
Addis Ababa, Ethiopia
(2009)
\end{botherref}
\endbibitem

\bibitem{AfricaEUCoop-Transport}
\begin{botherref}
\oauthor{\bparticle{for} \bsnm{Mobility}, \binits{E.C.-D.-G.}},
\oauthor{\bsnm{Transport}}:
Towards an enhanced {A}frica-{EU} cooperation on transport and connectivity.
Technical report,
European Commission
(2020)
\end{botherref}
\endbibitem

\bibitem{prieto2020uncovering}
\begin{barticle}
\bauthor{\bsnm{{Prieto-Curiel}}, \binits{R.}},
\bauthor{\bsnm{{Walther}}, \binits{O.}},
\bauthor{\bsnm{{O’Clery}}, \binits{N.}}:
\batitle{Uncovering the internal structure of {B}oko {H}aram through its mobility patterns}.
\bjtitle{Applied Network Science}
\bvolume{5}(\bissue{1}),
\bfpage{1}--\blpage{23}
(\byear{2020})
\end{barticle}
\endbibitem

\bibitem{bruneau2003framework}
\begin{barticle}
\bauthor{\bsnm{{Bruneau}}, \binits{M.}},
\bauthor{\bsnm{{Chang}}, \binits{S.E.}},
\bauthor{\bsnm{{Eguchi}}, \binits{R.T.}},
\bauthor{\bsnm{{Lee}}, \binits{G.C.}},
\bauthor{\bsnm{{O'Rourke}}, \binits{T.D.}},
\bauthor{\bsnm{{Reinhorn}}, \binits{A.M.}},
\bauthor{\bsnm{{Shinozuka}}, \binits{M.}},
\bauthor{\bsnm{{Tierney}}, \binits{K.}},
\bauthor{\bsnm{{Wallace}}, \binits{W.A.}},
\bauthor{\bsnm{{Von Winterfeldt}}, \binits{D.}}:
\batitle{A framework to quantitatively assess and enhance the seismic resilience of communities}.
\bjtitle{Earthquake Spectra}
\bvolume{19}(\bissue{4}),
\bfpage{733}--\blpage{752}
(\byear{2003})
\end{barticle}
\endbibitem

\end{thebibliography}

\end{backmatter}

\end{document}